\documentclass[aps,prb,twocolumn,letterpaper]{revtex4}
\usepackage{makeidx}
\usepackage{amsmath,amssymb,amsfonts,amsthm}
\usepackage{graphicx,bm}

\begin{document}

\title{Hopping and clustering of oxygen vacancies in SrTiO$_{3}$ by
anelastic relaxation}
\date{}
\author{F. Cordero}
\affiliation{$^1$ CNR-ISC, Istituto dei Sistemi Complessi, Area della Ricerca di Roma -
Tor Vergata,\\
Via del Fosso del Cavaliere 100, I-00133 Roma, Italy}

\begin{abstract}
The complex elastic compliance $s_{11}\left( \omega ,T\right) $ of SrTiO$%
_{3-\delta }$ has been measured as a function of the O deficiency $\delta
<0.01$. The two main relaxation peaks in the absorption are identified with
hopping of isolated O vacancies over a barrier of 0.60~eV and reorientation
of pairs of vacancies involving a barrier of 1~eV. The pair binding energy
is $\simeq 0.2$~eV and indications for additional clustering, possibly into
chains, is found already at $\delta \sim 0.004$. The anistropic component of
the elastic dipole of an O\ vacancy is $\Delta \lambda =0.026$.
\end{abstract}

\pacs{61.72.Ji,66.30.Dn,62.40.+i}
\maketitle


Diffusion and aggregation of oxygen vacancies (V$_{\text{O}}$) in
perovskites are still poorly understood, although they have implications in
several types of applications, like in solid state electrolytes for fuel
cells or fatigue in ferroelectrics. One of the most studied perovskites is
SrTiO$_{3-\delta }$, which has been heavily doped with V$_{\text{O}}$ since
the 1960s \textit{e.g. }for inducing superconductivity, but only recent
experiments [\cite{SSC02,SJR03}] raised questions on the assumption that the
V$_{\text{O}}$ introduced by high temperature reduction are uniformly
distributed over the bulk. In addition, there is a large spread of the
estimates of the activation energy for their diffusion [\cite{PKR99}],
although the prevalent opinion is that in most titanate perovskites the
barrier for V$_{\text{O}}$ hopping is $\simeq 1$~eV [\cite{WM05,CLM00,120}].
An O vacancy in a perovskite lattice has tetragonal symmetry and therefore
an associated tetragonal distortion, or elastic dipole, which reorients by 90%
$^{\mathrm{o}}$ after each jump causing anelastic relaxation
[\cite{NB72}]; in addition, being a charged defect, a hopping
V$_{\text{O}}$ causes fluctuations of the electric field that,
thanks to the quadrupolar interaction with the surrounding nuclei,
may be detected by nuclear magnetic relaxation (NMR). Indeed, two
elaxation processes are found by NMR in SrTiO$_{3}$, with activation
energies of 0.62~eV and 1.1~eV and attributed to hopping of
V$_{\text{O}}$ which are free or trapped by Fe, respectively
[\cite{HK91}]. On the other hand, SrTiO$_{3}$ reduced in
H$_{2}$ presents a rich anelastic relaxation spectrum, including
two processes with 0.6 and $\simeq 1$~eV, tentatively attributed to H and V$%
_{\text{O}}$ hopping [\cite{120}]. Here anelastic relaxation measurements on
a SrTiO$_{3}$ crystal reduced in CO are presented, providing evidence that
the barrier for hopping of isolated vacancies is indeed 0.60~eV, as also
estimated in recent calculations [\cite{CLC07}], but aggregation into pairs
and possibly chains occurs already at low $\delta $, so explaining the
higher activation energy for diffusion generally found.

The sample was cut from a SrTiO$_{3}$ wafer from M.T.I. Corporation as a bar
$26.15\times 3.4\times 0.5$~mm$^{3}$ with the edges parallel to the $%
\left\langle 100\right\rangle $ directions. The sample was covered with
silver paint on a face, suspended on thin thermocouple wires at the nodal
lines and electrostatically excited on its first and fifth flexural modes at
5.5 and 74~kHz; the electrode-sample capacitance was part of a resonant
circuit, whose frequency ($\sim 10$~MHz) was modulated by the sample
vibration, so allowing the vibration to be detected. The real part of the $%
s_{11}$ compliance could be measured from the fundamental resonance
frequency $f$ as $s_{11}^{\prime }=\rho ^{-1}\left( 0.973fl^{2}/h\right)
^{-2}=$ $3.64\times 10^{-13}$~cm$^{2}$/dyn at room temperature, where $l$, $%
h $ and $\rho $ are sample length, thickness and density [\cite{NB72}]. The
elastic energy loss coefficient $Q^{-1}=s_{11}^{\prime \prime
}/s_{11}^{\prime }$ was measured from the free decay of the sample vibration
or from the width of the resonance curve. The jumps of V$_{\text{O}}$ cause
a reorientation by 90$^{\mathrm{o}}$ of the direction of the nearest
neighbor Ti atoms and therefore also of the associated tetragonal elastic
dipole $\lambda $ (actually a quadrupole), having only two independent
diagonal elements $\lambda _{1}$ and $\lambda _{2}$; the resulting
relaxation $\delta s_{11}^{\prime \prime }$ of the compliance causes a peak
in $Q^{-1}\left( \omega ,T\right) $ [\cite{NB72}]
\begin{equation}
Q^{-1}=\frac{\delta s_{11}^{\prime \prime }}{s_{11}^{\prime }}=\frac{2}{9}%
\frac{cv_{0}}{s_{11}^{\prime }k_{\text{B}}T}\left( \Delta \lambda \right)
^{2}\frac{\alpha \left( \omega \tau \right) ^{\alpha }}{1+\left( \omega \tau
\right) ^{2\alpha }}  \label{peak}
\end{equation}%
where $\Delta \lambda =\lambda _{1}-\lambda _{2}$, $c$ is the concentration
of the relaxing defect, $v_{0}$ the molecular volume, $\omega =2\pi f$, $%
\tau =\tau _{0}e^{W/k_{\text{B}}T}$ is the relaxation rate over a barrier $W$%
, the parameter $\alpha \leq 1$ reproduces a possible broadening and the
maximum occurs at the temperature where $\omega \tau =1$.

For the reduction/oxidation treatments the sample was inserted in a
flattened cylindrical holder of Pt that was exposed to a flux of 1~bar of O$%
_{2}$ or 0.9 Ar + 0.1 CO and heated by induction within a quartz tube cooled
with water. Each reduction was preceded by oxygenation for 1.5~h at 950~$^{%
\mathrm{o}}$C that restored the white translucent aspect. The O\ deficiency
was deduced from the mass change and from the temperature $T_{0}$ of the
cubic-to-tetragonal transformation appearing as a step in the elastic
compliance near 105~K [\cite{120}]. The highest deficiencies after reducing
treatments of 3~h at 1100 and 1150~$^{\mathrm{o}}$C were estimated from the
mass loss as $0.0066\pm 0.001$ and $0.0070\pm 0.001$ and resulted in $%
T_{0}=92.4$ and 91.7~K respectively. The assumption of a linear $T_{0}\left(
\delta \right) $ combined with $T_{0}\left( 0\right) =105.2$ and 106.9~K
before and after the cycle of experiments yield $T_{0}\left( \delta \right) =
$ $\left( 106-2050~\delta \right) $~K; the reported doping $\delta $ is
deduced from this relationship. The shift of $T_{0}\left( 0\right) $
suggests that the 14 high temperature treatments for reduction,
homogenization and reoxidation, each preceded by mechanical removal of the
silver paste, may have introduced some damage in the sample (only part of
the results is reported here).

\begin{figure}
  \includegraphics[width=8.5 cm, trim = 0 0.5cm 0 0]{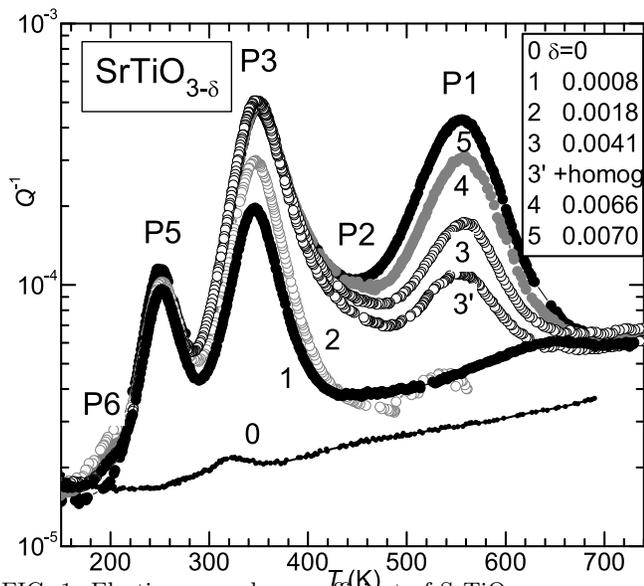}
  \caption{Elastic energy loss coefficient
of SrTiO$_{3-\protect\delta }$ measured at 5.4 kHz at various concentrations
of V$_{\text{O}}$. The sequence is according to $\protect\delta $ and not
chronological.}
  \label{fig allf1}
\end{figure}

Figure \ref{fig allf1} presents the $Q^{-1}\left( T\right) $ curves in the
as-received state (curve 0), and after various reduction treatments: $\delta
=0.00088$ after 1~h at 950~$^{\mathrm{o}}$C (curve 1), $\delta =0.0018$
after 1~h at 1000~$^{\mathrm{o}}$C (curve 2), $\delta =0.0041$ after 3~h at
1085~$^{\mathrm{o}}$C (curve 3), $\delta =0.0066$ after 3~h at 1100~$^{%
\mathrm{o}}$C (curve 4), $\delta =0.0070$ after 3~h at 1150~$^{\mathrm{o}}$C
(curve 5). After reduction, the sample was cooled to room temperature in 2
min (curves 3,4 and 5) or homogenized for 1~h in the same reducing
atmosphere at 800~$^{\mathrm{o}}$C (curves 1, 2, $3^{\prime }$); at such low
temperature no further O loss occurs. The $Q^{-1}\left( T\right) $ curves
confirm our previous measurements on ceramic samples [\cite{120}], with 6
peaks here labeled P1-P6 starting from high temperature.

All of them are shifted to higher temperature when measured at
higher frequency, indicating that they are due to thermally activated
relaxation processes; this is shown for curves 1, 2 and 5 in Fig. \ref{fig
f1f5}. Peak P1 has an activation energy of $\simeq 1$~eV and therefore it
had been associated with the jumps of V$_{\text{O}}$(we were not aware of
Ref. [\cite{HK91}]), while peak P3 had been associated with H hopping, since
its intensity saturates immediately and the reductions were carried out in H$%
_{2}$ atmosphere [\cite{120}]. Here there is no reason for H contamination,
and therefore such an assignment must be excluded. Peak P5 is certainly
associated with doping, but with a very weak dependence for $\delta >0.001$;
it might be related to polaronic relaxation and it will not be considered
further; a minor peak P6 and indications for another peak between P3 and P5
are hardly detectable.

\begin{figure}
  \includegraphics[width=8.5 cm, trim = 0 0.5cm 0 0]{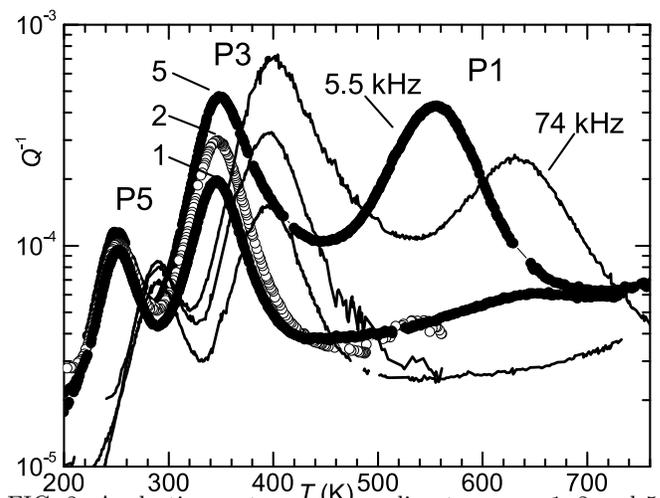}
  \caption{Anelastic spectra corresponding
to curves 1, 2 and 5 of Fig. 1, measured on the 1st (circles) and 5th
(lines) modes.}
  \label{fig f1f5}
\end{figure}

The frequency dependence of the anelastic spectra is shown for curves 1, 2
and 5 in Fig. \ref{fig f1f5}, and the changes in the intensities of the
peaks at higher frequency are fully meaningful, since both frequencies were
measured during the same run. The main features are: \textit{i)} the
intensity of P3 is $\propto 1/T$ at the lowest $\delta $, when P1 is
undetectable, and \textit{ii)} becomes a strongly increasing function of $T$
at higher $\delta $; \textit{iii)} the intensity of P1 decreases faster than
$1/T$. All these observations may be explained if P3 is due to hopping of
isolated V$_{\text{O}}$, while P1 is due to the reorientation of pairs of V$%
_{\text{O}}$. Indeed, the activation energy of P3, 0.60~eV, is in good
agreement with the NMR experiment [\cite{HK91}], while clustering of the V$_{%
\text{O}}$ in SrTiO$_{3}$ has been proposed [\cite{SSC02,SJR03}] and recent
calculations suggest that 2nd neighbor pairs of V$_{\text{O}}$ should be
stable [\cite{CLC07}]. The proposed picture is as follows: at very low O
deficiency, practically all V$_{\text{O}}$ are free, with a concentration $%
c_{f}\simeq \delta $; the intensity of P1 is $\propto c_{f}/T$ and therefore
$\propto 1/T$; already at $\delta \geq 0.001$ the concentration $c_{p}$ of
pairs of V$_{\text{O}}$ is significant and, due to thermal dissociation, $%
c_{p}\left( T\right) $ is a decreasing function, while $c_{f}\left( T\right)
$ becomes increasing, since $c_{f}+2c_{p}\simeq \delta $. This explains both
the pronounced decrease of the height of P1 with $T$ and the fact that the
height of P3 reverts to an increasing function of $T$. In view of these
considerations, P5 could not be assigned to hopping of isolated V$_{\text{O}}
$ because its intensity always decreases with temperature, whereas it should
increase due to the thermal dissociation of V$_{\text{O}}$ pairs, when P1 is
present.

\begin{figure}
  \includegraphics[width=8.1 cm, trim = 0 -0.5cm 0 0]{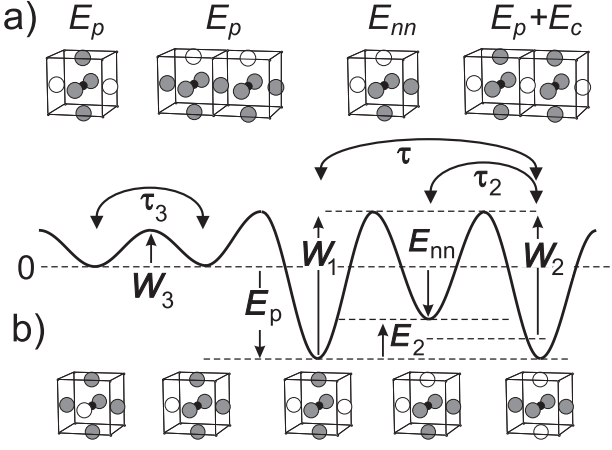}
  \caption{a) possible configurations of
two or three V$_{\text{O}}$ (empty circles) with the respective energies.
The Ti atoms are at the cell centres and Sr at the vertices. b) potential
energy profile for various jumps of V$_{\text{O}}$; below each minimum is
shown the corresponding configuration.}
  \label{fig pot_conf}
\end{figure}

The first attempts to fit all the spectra have been done taking into account
only isolated and paired V$_{\text{O}}$, whose concentrations may be
estimated as stationary solutions of rate equations for the formation of
pairs with binding energy $E_{p}$. In this manner, however, it is impossible
to fit all the spectra with the same set of $E_{p}$ and elastic dipole
anisotropies $\Delta \lambda $; in fact, the actual intensity of P1
increases with $\delta $ much less than predicted by the model. This
suggests further clustering of V$_{\text{O}}$, \textit{e.g.} formation of
chains along the $\left\langle 100\right\rangle $ directions, which are
predicted to be stable [\cite{CLC07}]; only the V$_{\text{O}}$ at the ends
would contribute to P1, since the internal V$_{\text{O}}$ would be bound
stronger and require a higher energy for jumping out of the chain, hence
contributing to anelastic relaxation at higher temperatures. The
quantitative description of the formation of V$_{\text{O}}$ chains cannot be
worked out in a simple manner, and it has been chosen to adopt the
grandcanonical formalism with the approximation of dividing the lattice into
small partitions, within which the formation of pairs and chains can be
treated exactly [\cite{33}]. Figure \ref{fig pot_conf}a) shows the
configurations of pairs and triplets of V$_{\text{O}}$ that have been
considered, besides the isolated V$_{\text{O}}$ with site energy $E_{f}=0$.
A binding energy $E_{p}$ is attributed to both pairs of the type V$_{\text{O}%
}-$Ti$-$V$_{\text{O}}$ and in adjacent cell faces; $E_{nn}$ to nearest
neighbor pairs and $E_{c}$ to each V$_{\text{O}}$ within a chain, while the
ends of a chain contribute with $E_{p}$.

The grandpartition function [\cite{Kit58}] $Z=\sum_{\alpha
}m_{\alpha }e^{\left( n_{\alpha }\mu -E_{\alpha }\right) /k_{\text{B}}T}$ is
written for three adjacent cells, so that it is possible to include the
formation of chains of up to four V$_{\text{O}}$. With three cells it is
still easy to count the multiplicities $m_{\alpha }$ of the possible
configurations $\alpha $ with $n_{\alpha }$ vacancies and with energy $%
E_{\alpha }$ by simple inspection, and the calculation of the chemical
potential $\mu $ and of the concentrations $c_{\alpha }$ is simple enough to
be integrated into the non-linear least square fitting routine. By defining $%
b=e^{-E_{p}/k_{\text{B}}T}$, $b_{c}=e^{-E_{c}/k_{\text{B}}T}$, $x=e^{\mu /k_{%
\text{B}}T}$, it results: $Z=$ $1+16x+$ $\left( 17b+67\right) x^{2}+$ $%
\left( 6b~b_{c}+88b\right) x^{3}+$ $\left( b~b_{c}^{2}+\text{ }%
24b^{2}\right) x^{4}$, where $16x$ is the statistical weight of a single V$_{%
\text{O}}$ in one of the 16 O sites of three cells, $17bx^{2}$ is the weight
of the possible 17 pairs with energy $E_{p}$, and so on. The nearest
neighbor configuration \textit{nn}, with energy $E_{nn}$ is the intermediate
step for the reorientation of a pair, but its statistical weight resulted
negligible and is omitted. The chemical potential $\mu $ must satisfy the
implicit equation
\begin{equation}
\delta =\frac{3}{16}c=\frac{3k_{\text{B}}T}{16}\frac{\partial \ln Z}{%
\partial \mu }=c_{f}+2c_{p}+c_{c}~,
\end{equation}%
where it is recognized that the maximum possible concentration $c=16$ over
three cells corresponds to $\delta =3$. This implicit equation for $x$ has
to be solved numerically for each $T$. The decomposition of $c$ into the
various $c_{\alpha }$ is obtained by keeping track of the contributions of
the various terms in $Z$: $c_{f}=$ $\frac{3}{16Z}\left(
16x+134x^{2}+88bx^{3}\right) $, $c_{p}=$ $\frac{3bx^{2}}{16Z}\left[ 17+\text{
}\left( 6b_{c}+88\right) x+\text{ }\left( b_{c}^{2}+48b\right) x^{2}\right] $
and $c_{c}=$ $\frac{3bb_{c}x^{3}}{8Z}\left( 3+b_{c}x\right) $.

In Fig. \ref{fig pot_conf}b) is shown the potential profile of the
relaxation processes corresponding to P1, P2 and P3, where $W_{i}$ are the
barrier heights and $E_{i}$ the site energies taking as zero the isolated V$%
_{\text{O}}$. Peak P3 is fitted with Eq. (\ref{peak}) with $c=$ $c_{f}$, and
P1 with $c=$ $c_{p}$; the intermediate relaxation involving the \textit{nn}
pair should have an intensity proportional to $c_{p}/[T\cosh ^{2}(E_{2}/2k_{%
\text{B}}T)]$, valid for the low concentration limit [\cite{33}], and rate $%
\tau _{2}^{-1}=\tau _{02}^{-1}e^{-W_{2}/k_{\text{B}}T}\cosh (E_{2}/2k_{\text{%
B}}T)$, where $E_{2}=2(W_{1}-W_{2})$ is the asymmetry between the two states.

\begin{figure}
  \includegraphics[width=8.5 cm, trim = 0 0.5cm 0 0]{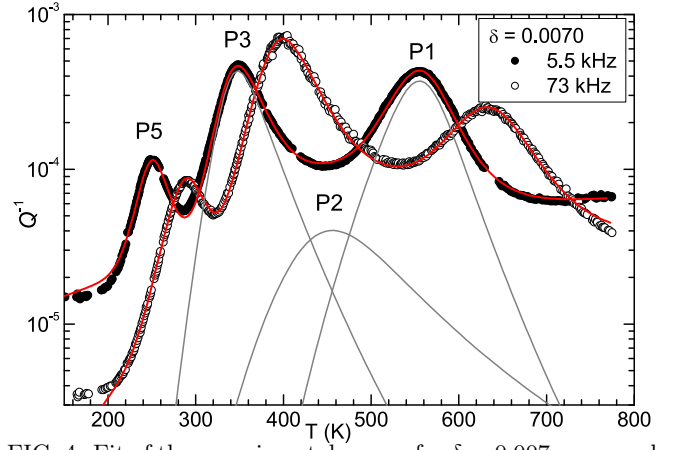}
  \caption{Fit of the experimental curves
for $\protect\delta =0.007$ measured at 5.5 and 73~kHz with $E_{p}=0.183$~eV
and $E_{c}=0.267$~eV; also shown are the components P1, P2 and P3 at 5.5~kHz.}
  \label{fig fitP63}
\end{figure}

The fit to the case $\delta =0.007$ is shown in Fig. \ref{fig fitP63} and
similar fits are obtained for the other concentrations $\delta >0.001$
assuming binding energies $E_{p}=0.184$~eV and $E_{c}=0.26$~eV. The
parameters for the isolated V$_{\text{O}}$ deduced from P3 are: anisotropic
component of the elastic quadrupole $\left( \Delta \lambda \right) _{3}=0.026
$, hopping rate with $W_{3}=0.60\pm 0.007$~eV and $\tau _{03}=\left( 5\pm
1\right) \times 10^{-14}$~s; for the pair reorientation (peak P1) $\left(
\Delta \lambda \right) _{1}=$ $1.87\times \left( \Delta \lambda \right) _{3}$%
, slightly less than for two independent V$_{\text{O}}$, and the
reorientation rate has $W_{1}=0.97\pm 0.04$~eV and $\tau _{01}=\left( 7\pm
4\right) \times 10^{-14}$~s; no broadening is found ($\alpha \geq 0.95$) for
P1 and P3. The intermediate relaxation P2 has $W_{2}\sim 0.86$~eV and $%
E_{2}\sim 0.17$~eV, but is much broader and more intense then expected from
the intermediate configuration of reorienting pairs, since $\alpha _{2}\sim
0.35$ and $\left( \Delta \lambda \right) _{2}\sim 5\left( \Delta \lambda
\right) _{3}$. It is therefore possible that also other configurations,
\textit{e.g.} third neighbor V$_{\text{O}}$, contribute to P2; recent
calculations indeed indicate that the interaction energy of V$_{\text{O}}$
may be quite large also at higher distances [\cite{CLC07}]. With $\Delta
\lambda \ $as small as $0.026$, the distortion around a V$_{\text{O}}$ is
almost isotropic (for interstitial O in \textit{bcc} metals is $\Delta
\lambda \sim 1$ [\cite{NB72}]), as predicted [\cite{LDL04}]. Peak P5 has $%
W_{5}=$ $0.43$~eV, $\tau _{05}=$ $\left( 1\pm 0.5\right) \times 10^{-13}$~s
and $\alpha _{5}=1$.

Only the spectrum with $\delta \simeq 8.8\times 10^{-4}$ (curve 1 of Fig. %
\ref{fig allf1}) cannot be reproduced with the same parameters, but requires
$\delta =6.4\times 10^{-4}$ , slightly smaller than estimated from the
transition temperature $T_{0}$ but within the error, and $E_{p}\simeq 0.1$
eV, otherwise it is impossible to obtain the $1/T$ dependence of the
intensity of P3. A possible explanation is that there is a fraction of V$_{%
\text{O}}$ that are strongly trapped at lattice defects; the influence of
such defects would be particularly evident in curve 1, since it has the
smallest $\delta $ and was also the last measurement of the series.
Indications of lattice damage accumulated during the treatments are the
presence in curve 1 of a broad peak at temperature higher than that of P1
and the shift of $T_{0}$ mentioned above.

Let us comment on the time for reaching true thermodynamic equilibrium. In
fact, after the sample was reduced at the highest values of $\delta $ and
cooled to room temperature within a couple of minutes, the distribution of V$%
_{\text{O}}$ was not in equilibrium. This is demonstrated by the effect of a
homogenization treatment of 1~h at 800~$^{\mathrm{o}}$C, here shown in
curves 3 and $3^{\prime }$ of Fig. \ref{fig allf1}: all the peaks remain
unaffected except for P1, which decreases. This certainly cannot be
explained by O uptake during homogenization, due to the absence of O$_{2}$\
in the reducing atmosphere, nor can it be explained in terms of further loss
of O, which occurs at higher temperatures and results in an increment of the
height of P1. It must be concluded that the concentration of V$_{\text{O}}$
pairs decreases during homogenization due to the formation of more stable O
clusters. Within the proposed picture, the suppression of P1 is explainable
in terms of lengthening of the chains of V$_{\text{O}}$ which reduces the
fraction of pairs and chain ends.

A seeming inconsistency exists between the fact that peaks P1, P2 and P3
imply hopping rates of isolated and aggregated V$_{\text{O}}$ exceeding 10$%
^{4}$~s$^{-1}$ already at 600~K, while the aggregation into longer chains
near 1100~K requires hours. The situation is quite similar to that of the
ordering of the O atoms in the CuO$_{x}$ planes of semiconducting YBa$_{2}$Cu%
$_{3}$O$_{6+x}$ [\cite{44}], which also are perovskite-like layers. In that
case, the isolated O atoms in the almost empty CuO$_{x}$ planes ($x<0.3$)
hop over a barrier as low as 0.11~eV and promptly form stable pairs or short
chains whose dissociation energy is $\sim 1$ eV, but reaching equilibrium
with longer O-Cu-O chains requires very long times. The explanation proposed
for such a behavior is that the saddle point for an O$^{2-}$ ion to join
another O$^{2-}$\ ion or chain fragment is higher than that for hopping away
from it, due to the electrostatic repulsion. The analogous effect is shown
for the V$_{\text{O}}$ in Fig. \ref{fig pot_conf}b), where the saddle points
involving pairs are higher than those between free V$_{\text{O}}$. If we
assume that all the saddle points involving pairs are similar, as depicted
in Fig. \ref{fig pot_conf}b), then they are higher than in the empty lattice
by $W_{1}-E_{p}-W_{3}\sim 0.19$~eV. Notice that the rate for forming and
dissociating a pair is not probed by any of the observed peaks (possibly by
P2), and it is possible that the enhancement of the corresponding saddle
point is even more than 0.19~eV. In this manner it would be explainable why
the rate for constructing and rearranging the chains is much slower than
that for the pair reorientation. The fact that it is difficult to assess
when the sample is in real thermodynamic equilibrium, together with the
approximate treatment of the lattice as an ensemble of triplets of cells,
put some limitation on the confidence of the binding energies $E_{p}$ and $%
E_{c}$ derived from the fits of the anelastic spectra.

In conclusion, the anelastic spectrum of SrTiO$_{3-\delta }$ exhibits
relaxation peaks that are assigned to  \textit{i)} hopping of isolated O
vacancies over a barrier of 0.60~eV,  \textit{ii)} reorientation of pairs of
vacancies over a barrier of 0.97~eV and  \textit{iii)} jumps involving the
intermediate step for the pair reorientation and other configurations.
Sizeable formation of pairs starts at $\delta $ as low as 0.002, and there
is evidence of further aggregation, possibly into chains. The pair binding
energy is estimated as 0.18~eV, while the binding energy of additional
vacancies in a chain is $\sim 0.26$~eV.

The author thanks P. Verardi for cutting the sample, F. Corvasce, M. Latino
and A. Morbidini for technical assistance, and acknowledges the financial
support of the FISR Project \textquotedblleft Celle a combustibile".


\end{document}